# Observation of anisotropic magneto-inductance effect

Yuto Shoka[1], Genki Okano[2], Hiroyuki Suto[2], Satoshi Sumi[1], Hiroyuki Awano[1], and Kenji Tanabe[1*]

[1]Toyota Technological Institute, Nagoya, Japan

[2]Toyota Motor Corporation, Toyota, Japan

*electronic mail: tanabe@toyota-ti.ac.jp

**Abstract**

We have discovered a new phenomenon that inductance oscillates as a function of the angle between an in-plane magnetic field and an electric current direction in permalloy films, which we have named "the anisotropic magneto-inductance (AML) effect". We have investigated the dependences of the AML effect on the size and voltage. The length, frequency, and amplitude dependences suggest that the AML effect should be evaluated in terms of "inductivity". Inductors based on this AML effect have the potential to be variable, on-chip, and one billion times smaller than the small commercial inductor.

A resistor, a capacitor, and an inductor are the most important passive components in the modern electronic circuit. The resistor and capacitor are based on material parameters such as resistivity and permittivity, and their research remarkably progresses with the development of material science. On the other hand, the shape of the inductor has not progressed from a solenoid coil, which had been developed over one hundred years ago.

Recently, Nagaosa has proposed a new model for inducing inductance, which does not require a solenoid structure[1]. It is derived from spin-motive force (SMF)[2-5], a new electromotive force induced by magnetic dynamics in non-uniform magnetic structures such as a magnetic domain wall[6,7], a comb structure[8,9], a magnetic vortex[10], a trapezoid structure[11,12], and a multi-layer structure[13]. Since the SMF is a quantum effect related to the s-d exchange interaction, the term "emergent inductor" is coined in contrast to solenoid inductors in classical electromagnetism. Ieda et al have examined the influence of the Rashba spin-orbit coupling (SOC) on emergent inductance and found that the emergent inductance is induced even in a uniform magnetic structure[14-15]. This emergent inductance can be controlled by the application of the electric field.

In 2020, Yokouchi et al. have demonstrated for the first time inducing emergent inductance in the spiral magnet $Gd_3Ru_4Al_{12}$[16]. Negative values of the inductance were detected in the temperature lower than 20 K, corresponding to the Neel point. A nonlinear response was observed in which the inductance depends on the current density. The inductance rapidly decreases above 10 kHz, which is possible to originate from the depinning of the spiral spin structure. Kitaori et al. have observed emergent electromagnetic induction in the spiral magnet $YMn_6Sn_6$ beyond room temperature[17]. Both the positive and negative signs of the emergent inductance were observed. After that, it has been clarified that the inductance can be controlled by substituting a part of Y by Tb in $Y_{1-x}Tb_xMn_6Sn_6$[18]. Matsushima et al. have reported inductance induced in NiFe films having a small in-plane magnetic anisotropy on plastic substrates[19]. The inductance was detected under the magnetic field lower than 10 Oe. They claimed that it originates from the SMF induced in the non-uniform magnetic states produced by the small in-plane magnetic anisotropy under the low magnetic field.

In this study, we have investigated the imaginary part of the anisotropic magneto-resistance (AMR) effect in permalloy films. The permalloy has a relatively large AMR ratio of almost 3% compared with other transition metals at room temperature [20-22]. We have found the oscillation of the inductance as a function of the angle between the magnetic field and electric current, which is named the anisotropic magneto-inductance (AML) effect. Since the AML effect has been detected under the magnetic field of 300 Oe, which is much larger than the coercive force, the origin is not the SMF. We have examined the dependences of the AML effect on the length, width, thickness of the sample, frequency, and amplitude of the ac voltage. It is clarified that the AML effect is evaluated by the inductivity. The inductor based on the AML effect has the potential to become one billion times smaller than the volume of a small conventional inductor, and an on-chip inductor embedded on a substrate. Since the inductance is modulated by the application of the magnetic field, our result may lead to realizing a variable inductor, which can modulate an inductance after the fabrication of the inductor.

The sample was fabricated by maskless photolithography (NEOARK corp.), a magnetron sputtering, and a liftoff method. $Si_3N_4$(10 nm)/$Ni_{80}Fe_{20}$($d$) was deposited on Si substrates having a thermally oxidized layer (350 nm). The $Si_3N_4$ layer acts as the protection layer for the oxidization of the permalloy layer. The coercive force along the in-plane direction is about 1 Oe, which was checked by a vibrating sample magnetometer. The sample structure is a strip having four pads as shown in Fig. 1(a). The length, width, and thickness of the strip are denoted by $l$, $w$, and $d$, respectively. A resistance and an inductance were measured using the LCR meters (Hewlett-Packard 4284a) in Fig. 1(b). The frequency is 10, 30, and 100 kHz, and the amplitude of the ac voltage is 0.1, 1 V. The angle is defined as 0 degree as the in-plane magnetic field is parallel to the current direction. An amplitude of the magnetic field is 300 Oe except for the experiment on the dependence on the magnetic field.

Firstly, we observed oscillation in the resistance $\Delta R$ as shown in Fig. 2(a). The oscillation of $\Delta R$ indicates a well-known AMR effect. The AMR ratio $\Delta R/R$ is almost $-1\%$. On the other hand, Figure 2(b) shows the angle dependence of inductance $\Delta L$. The oscillation of $\Delta L$ was observed for the first time and named an AML effect. This inductance increases with a decrease in resistance and decreases with an increase in resistance. The AML ratio $\Delta L/L$ is almost $-1.7\%$. We have examined the length $l$ dependence of the AML effect in order to characterize the AML effect. The AML effect increases with increasing $l$ as shown in Fig. 2(b), which is quite similar to the trend of the change in resistance. Here, the inductance, which is an extensive variable, is converted to an inductivity, which has been proposed as an intensive variable in the previous report[23], as below.

$$\Delta\iota = \Delta L \frac{wd}{l}. \qquad (1)$$

The inductivity is not a parameter to evaluate a conventional solenoid inductor unlike permeability. Figures 2(c-d) show the conversion results from the resistance and inductance to the resistivity $\Delta\rho$ and inductivity $\Delta\iota$, respectively. Needless to say, the resistivity is independent of the length of the sample. The inductivity $\Delta\iota$ is also independent of the length, which reveals that the inductance can be evaluated with an intensive variable.

Secondly, we investigated the width and thickness dependencies of the AML effect. Figures 3(a-b) show the width $w$ dependences of $\Delta\rho$ and $\Delta\iota$, and Figures 3(c-d) show the thickness $d$ dependences of $\Delta\rho$ and $\Delta\iota$. $\Delta\rho$ is independent of $w$ and $d$ in the region thicker than 30 nm, and slightly increases in the region thinner than 10 nm. The increase of $\Delta\rho$ in the thin region stems from the fact that the thickness is comparable to the mean free path of electrons, which promotes the interface scattering of conduction electrons. On the other hand, $\Delta\iota$ strongly depends on both $w$ and $d$, and $\Delta\iota$ becomes larger in thinner and narrower strips. The inductivity is not an intensive variable in this study. The results mean that $\Delta\iota$ is not directly combined with $\Delta\rho$ and that the Q factor, which is defined as $Q = \omega L/R (= \omega\iota/\rho)$, can be improved by optimization of the width and thickness. Here, we compare the size of a realizable inductor based on the AML effect with the small conventional inductor. The commercial chip inductors are 0.8 mm long, 0.4 mm wide, and almost 0.4 mm thick in the case of the small inductors (LQP series) produced by Murata corp. Since it is easy to fabricate an inductor having a length of 3 μm, a width of 3 μm, and a thickness of 10 nm by standard photolithography, the inductor can become almost one billion times smaller than the volume of the commercial inductors.

Thirdly, we investigated the dependences on the frequency, the voltage, and the amplitude of the magnetic field. Figures 4(a) and 4(b) show the inductivity as functions of the frequency and the voltage, respectively. The inductivity slightly decreases with increasing frequency and is independent of the voltage at 0.1 and 1 V. Although the emergent inductance detected in the previous studies has a strong nonlinearity in that the inductance depends on the current density[16-18], the AML effect shows the linear response, which is an advantage for commercial applications. The inductance can be evaluated as not only the inductivity but also the imaginary part of the ac resistivity, which is defined as a product of the angular frequency and inductivity, $\omega\iota$. Indeed, Kitaori et al. have analyzed the emergent inductance in the spiral magnet using the imaginary part of the resistivity[17-18]. Since the AML effect, however, is almost independent of the frequency, it is clarified that it should be evaluated as the inductivity rather than as the imaginary part of the ac resistivity. Moreover, the fact that the inductance is independent of the frequency is important for commercial applications. Figures 4(c) and 4(d) show the resistance and inductance as a function of the magnetic field, respectively. It is noted that a static magnetic field of 300 Oe was used for the experiments in Figs. 2, 3, 4(a-b), and 5 and that the magnetic field was varied only for the experiments in Figs. 4(c-d). A magnetoresistance effect occurs when the magnetic field is small. It stems from the AMR effect in a multi-domain state. At the same time, the change in inductance also occurs at the small magnetic field. The resistance increases and the inductance decreases as the magnetic field is parallel to the current direction. On the other hand, the resistance decreases and the inductance increases as the magnetic field is perpendicular to the current direction. These tendencies completely accord with the AML effect and are possible to be not directly related to non-uniform magnetic structures such as a multi-domain state but perpendicular component of the magnetization to the current.

Fourthly, we consider the influence of the sample geometry on the AML effect. The AML effect is enhanced with the thinner and narrower strip. The composition structure that consists of thinner and narrower strips, therefore, is expected to induce a larger AML effect. We fabricated a simple sample consisting of one wide strip and a complex sample consisting of three narrow strips as shown in Figs. 5(a-b). The total width is the same. If the AML effect in the complex sample followed the results on the width dependence in Fig. 3(b), the complex sample would induce a larger AML effect. Surprisingly, both of the results merge in one curve. Moreover, we fabricated a thick monolayer and a multilayer consisting of three thin layers as shown in Fig. 5(d-e). The total thickness is the same. The data in the multilayer also accords with that in the monolayer. These results are inconsistent with the width- and thickness-dependent results, meaning that a long-range interaction between the strips is important for the AML effect.

Finally, let us discuss the origin of the AML effect. The first candidate is the emergent inductance based on the SMF Nagaosa proposed[1]. Matsushima et al. have reported the appearance of inductance in the permalloy films having the weak in-plane magnetic anisotropy and have claimed that the SMF is induced by the slight distortion of the magnetic structure[19]. The AML effect, however, is observed under a magnetic field of 300 Oe, which is considerably quite larger than the coercive force, and the magnetic structure is completely a single magnetic domain. Since the SMF is not induced in a single magnetic domain but in non-uniform magnetic structures, the AML effect does not originate from the SMF. The second candidate is the emergent inductance on the Rashba SOC that Idea et al. studied[14-15]. The Rashba SOC does not require the non-uniform magnetic structure[24]. It has been believed that the Rashba SOC is induced in asymmetrical structures such as AlO$_x$/ultrathin Co/Pt[25]. According to this conventional understanding, the thicknesses of our permalloy films are too thick to induce the Rashba SOC. The theory using the Rashba SOC cannot explain the AML effect. However, recently, it has been reported that some SOCs such as the Dzyaloshinskii–Moriya interaction and inverse spin Hall effect work in ferrimagnetic layers with perpendicular composition gradients and in Py monolayers having symmetrical structures[26-28]. Some SOCs might work in our Py films and further research is needed. The third candidate is the skin depth effect in the radio-frequency region, which is a famous phenomenon as the origin of the high-sensitive magnetic impedance sensor[29-34]. The skin depth effect is strongly enhanced in the high-frequency regions. The AML effect, however, slightly decreases with an increase in frequency and is detectable in the relatively low-frequency regions. Moreover, the skin depth effect is related to permeability and is suppressed by a large magnetic field that pins magnetic moments. The AML effect is almost independent of the amplitude of the magnetic field of more than 5 Oe. The AML effect, therefore, cannot be expressed by the skin depth effect. The fourth candidate is inductance based on nonlinear conduction. Tanabe et al. have reported the appearance of the giant inductance in $Ca_2RuO_4$, which has non-ohmic conduction[23]. The inductance requires a large decrease in resistance due to current. In the permalloy films, the change in resistance due to current occurs owing to the AMR effect. The AMR ratio, however, is a few percent in the permalloy films[20-22], which is too small to induce the inductance based on the non-ohmic conduction. Therefore, although the origin of the AML effect is not clarified, it may be an important key to solving the origin that the AML effect has curious characteristics on the dependences of the inductivity on the width and thickness.

In summary, we have discovered the AML effect in the thin permalloy films, which indicates the oscillation of the inductance as a function of the angle between an in-plane magnetic field and an electric current direction. We have investigated the dependences of the AML effect on the length, width, thickness of the sample, frequency, and amplitude of the ac voltage. The length, frequency, and amplitude dependences suggest that the AML effect should be evaluated in terms of inductivity. The width and thickness dependences reveal that the inductance becomes larger in the thinner and wider strip. The inductor based on the AML effect has the potential to become one billion times smaller than the volume of a small conventional inductor, and an on-chip inductor embedded on a substrate. Since the inductance is modulated by the application of the magnetic field, our result may lead to realizing a variable inductor, which can modulate an inductance after the fabrication of the inductor.

**Acknowledgement**

This work was partially supported by a Grant-in-Aid for Scientific Research (C) (No. 20K05307) from JSPS.

**Figure 1**

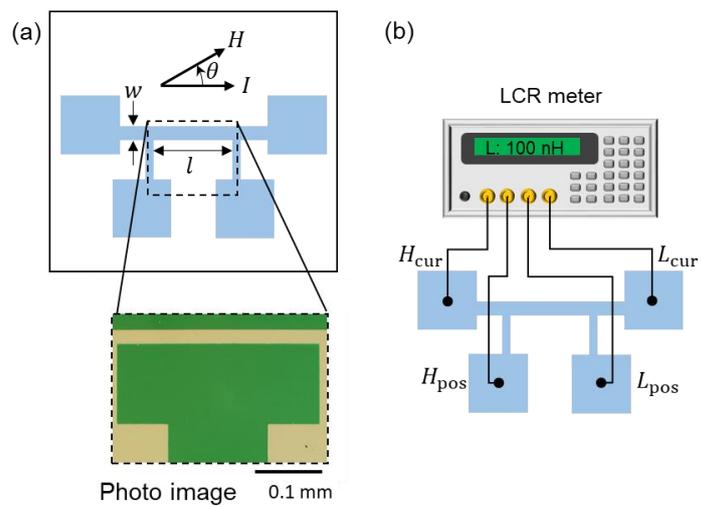

Fig. 1. (a-b) Schematic illustrations of sample structure(a) and measurement setup(b). The figure below in Fig. 1(a) is the optical microscope image of the sample.

**Figure 2**

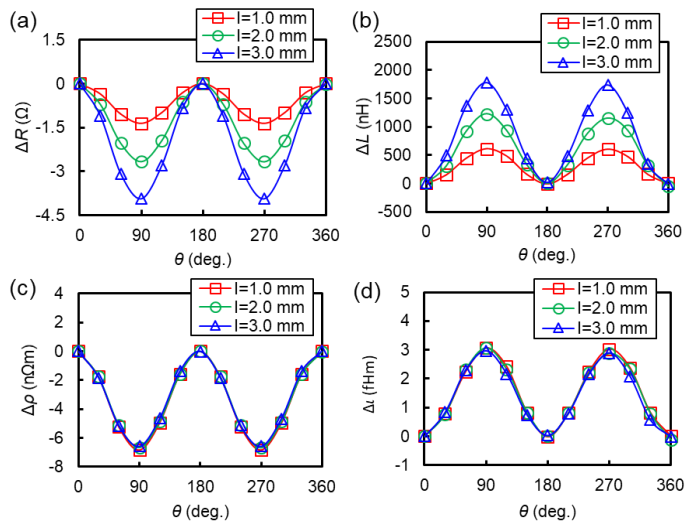

Fig. 2. (a-b) The resistance(a) and inductance(b) as functions of the field angle $\theta$ and the sample length $l$. $\Delta R (\Delta L)$ is defined as the difference from the resistance (inductance) to the one at $\theta$ = 0 deg. An amplitude of a static magnetic field is 300 Oe. The sample length is varied from 1.0 to 3.0 mm. The square, circle, and triangle indicate $l = 1.0$ mm, $l = 2.0$ mm, and $l = 3.0$ mm, respectively. $w = 0.1$ mm, $d = 50$ nm, $f = 10$ kHz, and $V = 1$ V. (c-d) The resistivity(c) and inductivity(d) as functions of the field angle $\theta$ and the sample length $l$. $\Delta\rho$ and $\Delta\iota$ are converted from $\Delta R$ and $\Delta L$ in Figs. 2(a) and 2(b), respectively.

**Figure 3**

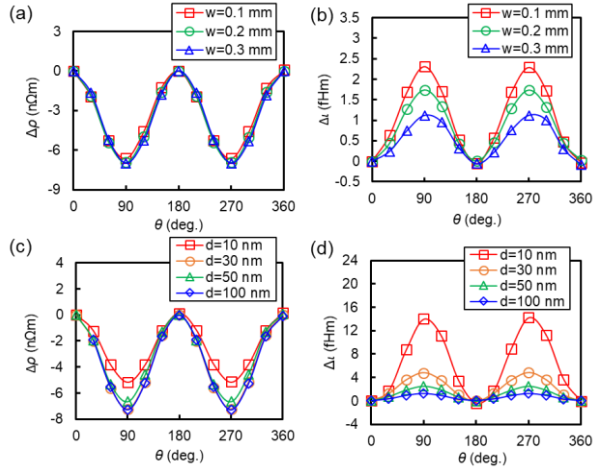

Fig. 3. (a-b) The resistivity(a) and inductivity(b) as functions of the field angle $\theta$ and the sample width $w$. The definitions of $\Delta\rho$ and $\Delta\iota$ are same as those in Fig. 2. An amplitude of a static magnetic field is 300 Oe. The sample width is varied from 0.1 to 0.3 mm. The square, circle, and triangle indicate $w = 0.1$ mm, $w = 0.2$ mm, and $w = 0.3$ mm, respectively. $l = 1.5$ mm, $d = 50$ nm, $f = 30$ kHz, and $V = 1$ $V$. (c-d) The resistivity(c) and inductivity(d) as functions of the field angle $\theta$ and the sample thickness $d$. The sample thickness is varied from 10 to 100 nm. The square, circle, triangle, and diamond indicate $d = 10$ nm, $d = 30$ nm, $d = 50$ nm, and $d = 100$ nm, respectively. $l = 1.5$ mm, $w = 0.1$ mm, $f = 10$ kHz, and $V = 1$ $V$.

**Figure 4**

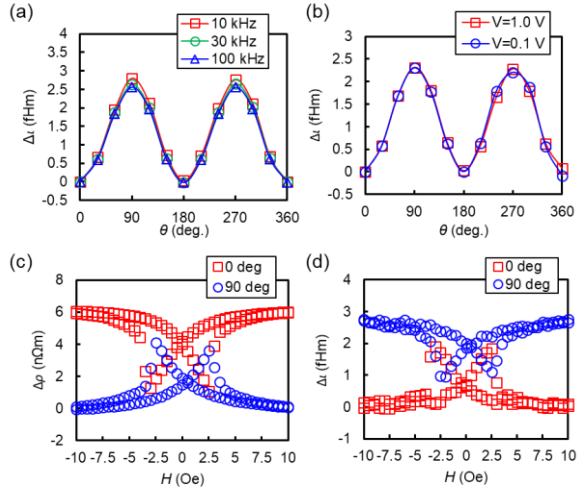

Fig. 4 (a) The inductivity as functions of the field angle $\theta$ and the frequency $f$. The definition of $\Delta\iota$ are same as that in Fig. 2. An amplitude of a static magnetic field is 300 Oe. The frequency is varied from 10 to 100 kHz. The square, circle, and triangle indicate $f = 10$ kHz, $f = 30$ kHz, and $f = 100$ kHz, respectively. $l = 1.5$ mm, $w = 0.2$ mm, $d = 30$ nm, and $V = 1$ V. (b) The inductivity as functions of the field angle $\theta$ and the voltage $V$. The definition of $\Delta\iota$ are same as that in Fig. 2. An amplitude of a static magnetic field is 300 Oe. The voltages are 1.0, 0.1 V. The square, and circle indicate $V = 1$ V and $V = 0.1$ V, respectively. $l = 1.5$ mm, $w = 0.1$ mm, $d = 50$ nm, and $f = 30$ kHz. (c-d) The dependences of $\Delta R$(c) and $\Delta L$(d) on the magnetic field. The definitions of $\Delta\rho$ and $\Delta\iota$ are same as those in Fig. 2. The circle and square indicate as the magnetic field is perpendicular and parallel to the current direction, respectively. $l = 1.5$ mm, $w = 0.1$ mm, $d = 50$ nm, $f = 10$ kHz, and $V = 1$ V.

**Figure 5**

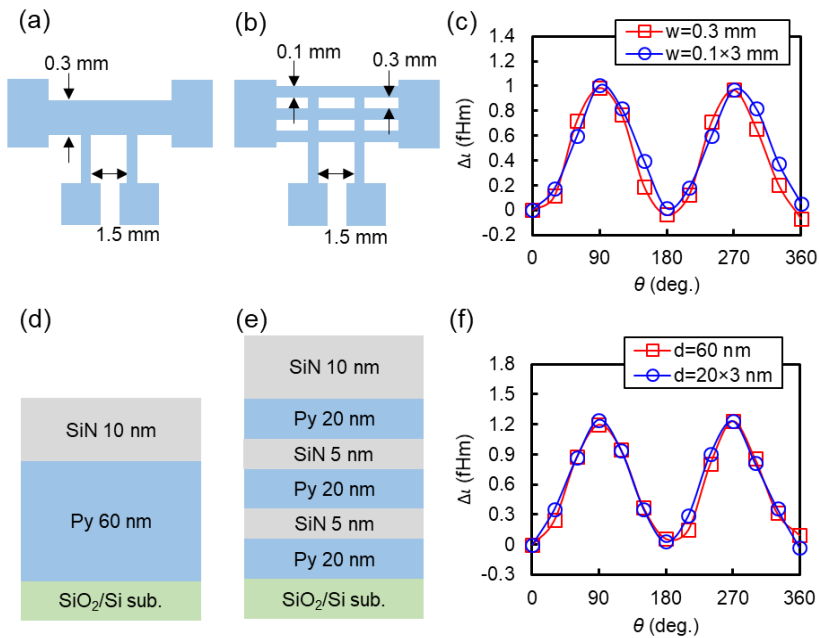

Fig. 5 (a-b) The schematic illustration of the sample structures. The samples are a strip with the width of 0.3 mm(a) and three strips with the width of 0.1 mm(b). The total widths are same. (c) The inductivity as functions of the field angle $\theta$ and the sample shape. The definition of $\Delta\iota$ are same as that in Fig. 2. An amplitude of a static magnetic field is 300 Oe. The square and circle indicate the one strip with the width of 0.3 mm and the three strips with the width of 0.1 mm, respectively. $l = 1.5$ mm, $d = 60$ nm, $f = 30$ kHz, and $V = 1$ V. (d-e) The schematic illustration of the stacking structures. The samples are a single layer with the thickness of 60 nm(d) and a multilayer that consists of three layers with the thickness of 20 nm(e). The total thicknesses are same. (f) The inductivity as functions of the field angle $\theta$ and the staking structure. The definition of $\Delta\iota$ are same as that in Fig. 2. An amplitude of a static magnetic field is 300 Oe. The square and circle indicate the single layer and multilayer, respectively. $l = 1.5$ mm, $w = 0.3$ mm, $f = 10$ kHz, and $V = 1$ V.